\documentclass[aps,titlepage,12pt,floatfix]{revtex4}
\usepackage{amsfonts}
\usepackage{amsmath}
\usepackage{graphicx}
\usepackage{booktabs}
\renewcommand{\arraystretch}{1.15}
\usepackage{color}
\usepackage[justification=raggedright,font={small,sf}, singlelinecheck=false]{caption}
\usepackage{hyperref}
\usepackage{url}

\graphicspath{{./}}

\begin{document}

\title{Demanding peer review is associated with higher impact in published science}

\author{Huihuang Jiang$^1$, Heyang Li$^2$, Zifan Wang$^3$, Ying Fan$^1$, and An Zeng$^{1,}$\footnote{Correspondence to: anzeng@bnu.edu.cn}}

\affiliation{
$^1$ School of System Science, Beijing Normal University, Beijing 100875, China.\\
$^2$ School of Public Policy and Management, University of Chinese Academy of Sciences, Beijing 100190, China.\\
$^3$ Business School, Beijing Normal University, Beijing 100875, China.
}

\begin{abstract}
Peer review shapes which scientific claims enter the published record, but its internal dynamics are hard to measure at scale because reviewer criticism and author revision are usually embedded in long, unstructured correspondence. Here we use a fixed-prompt large language model pipeline to convert the review correspondence of \textit{Nature Communications} papers published from 2017 to 2024 into structured reviewer--author interactions. We find that review pressure is concentrated in the first round and focused disproportionately on core claims rather than peripheral presentation. Higher average opinion strength is also associated with more reviewer disagreement, while review patterns vary little with broad team attributes, consistent with relatively impartial evaluation. Contrary to the intuition that stronger papers should pass review more smoothly, with greater reviewer--author agreement and less extensive revision, we find that stronger criticism, higher-quality comments, and greater revision burden are associated with higher later citation impact within accepted papers. We finally show that fields differ more in review style than in review length, pointing to disciplinary variation in how criticism is negotiated and resolved. These findings position open peer review not just as a gatekeeping mechanism but as a measurable record of how influential scientific claims are challenged, defended, and revised before entering the published record.
\end{abstract}

\date{\today}

\maketitle

\section{Introduction}

The science of science seeks to explain how knowledge is produced, evaluated, and rewarded using large-scale data and computational analysis \cite{fortunato2018science}. Peer review lies at the centre of this system. It is the main institutional process through which manuscripts are scrutinized before publication, claims are challenged, and scientific credibility is provisionally assigned \cite{bornmann2008effectiveness,aczel2025future}. Yet the system is under growing strain. The expansion of global scientific output has intensified editorial workload, reviewer recruitment has become more difficult in some fields, and review labour is increasingly unevenly distributed \cite{hanson2024strain,fox2017recruitment,severin2021overburdening}. These pressures coincide with broader distortions in academic evaluation, including hyper-prolific publishing and the over-optimization of publication metrics \cite{ioannidis2018thousands,fire2019goodhart,biagioli2020gaming}. Understanding peer review is therefore essential not only because it filters manuscripts, but also because it helps govern how scientific claims are challenged, revised, and admitted into the published record.

A large literature has examined peer review from several complementary angles. One recurring finding is that reviewers evaluating the same manuscript or grant proposal often diverge substantially in their judgments \cite{bornmann2010reliability,mutz2012heterogeneity,pier2018lowagreement}. Another is that peer review can be shaped by gender, prestige, institutional affiliation, race, ethnicity, and other status signals, which has sustained interest in reforms such as double-blind review \cite{lee2013bias,vanderlee2015gender,witteman2019gender,squazzoni2021gendergap,murray2021genderbias,ginther2011race,blank1991doubleblind,tomkins2017singleblind,kern2022doubleblind}. At the same time, scholars have raised the concern that peer review may be less hospitable to highly novel, unconventional, or high-variance ideas \cite{wang2017noveltybias,boudreau2016novelty,lee2015commensuration,shaw2022pursuitworthiness}. Journals and publishers have also experimented with open peer review and other innovations intended to increase transparency and accountability \cite{kaltenbrunner2022innovating,bravo2019reports}. Together, these literatures show that peer review is important and imperfect, but they leave unresolved what intense review actually means within papers that eventually get published.

The main reason is that most existing analyses remain at the level of whole reports, whole manuscripts, or final decisions rather than the level of discrete issues exchanged between reviewers and authors. Large-scale studies of review text have shown systematic variation in sentiment, style, moral language, and length \cite{buljan2020language}, and related work has argued for broader peer-review data sharing as a foundation for cumulative metaresearch \cite{squazzoni2020unlockways}. Recent Natural language processing (NLP) research has also created benchmark resources for peer-review text and surveyed how computational methods can assist review analysis \cite{ghosal2022peerreviewanalyze,checco2021aiassistedpeerreview,kuznetsov2024nlppeerreview}. A smaller literature asks whether editorial selection and peer review are related to later citation outcomes, but again at the level of acceptance decisions or manuscript trajectories rather than issue-level interaction \cite{bornmann2008selectingmanuscripts}. What remains comparatively underdeveloped is a structured account of reviewer--author exchange across rounds: who raised which concern, how authors responded, whether the issue persisted, and what kinds of criticism accompany later impact.

Large language models (LLMs) make that issue-level measurement problem more tractable. Recent work shows that LLMs can perform demanding text-annotation tasks at high quality and may transform measurement workflows in computational social science when task design and validation are explicit \cite{gilardi2023chatgptannotators,ziems2024css,tornberg2024bestpractices}. For peer review, the opportunity is methodological rather than promotional. LLMs can be used to segment long correspondence into issue-specific units, infer links between reviewer comments and author replies, and recover comparable measurements from review text that would otherwise remain too heterogeneous for large-scale analysis.

Here we use that opportunity to analyse the full-cycle peer review correspondence of a year-balanced sample of 8,000 \textit{Nature Communications} papers published between 2017 and 2024. We use a fixed-prompt large language model procedure to convert the correspondence into LLM-extracted comment--response pairs and measure seven dimensions of interaction: opinion strength, constructiveness, comment quality, revision cost, opinion type, author response type, and persuasion success. Throughout the paper, we use \emph{demanding review} as a descriptive shorthand for combinations of stronger criticism, higher comment quality, and greater revision cost.

Our central question is what kinds of review profiles are associated with higher later impact among published papers. That question matters for editors, authors, and science policy because it helps identify how scrutiny is allocated within the published record. We show that demanding review is concentrated in the first round and directed disproportionately at core claims rather than peripheral presentation. We further show that, among papers reviewed by three first-round reviewers, dissenter configurations become more common as average opinion strength rises and rebuttal becomes more likely as the number of dissenters increases. Broad team attributes still explain relatively little overall variation, consistent with a relatively impartial review process, while cross-field differences are more visible in review style than in review length, suggesting disciplinary variation in how criticism and disagreement are negotiated. These results also challenge the common intuition that the highest-quality papers should move most smoothly through review. We find instead that papers that attract deeper scrutiny, more disagreement over central claims, and more substantial revision are often the ones with greater later impact. Together, these findings suggest that peer review is not simply a gatekeeping filter but a measurable interactional process through which influential scientific claims are challenged, negotiated, and refined.

\section{Results}

\subsection{Review measures are reproducible across models and validation checks}

\begin{figure}[p]
  \centering
  \includegraphics[width=\textwidth]{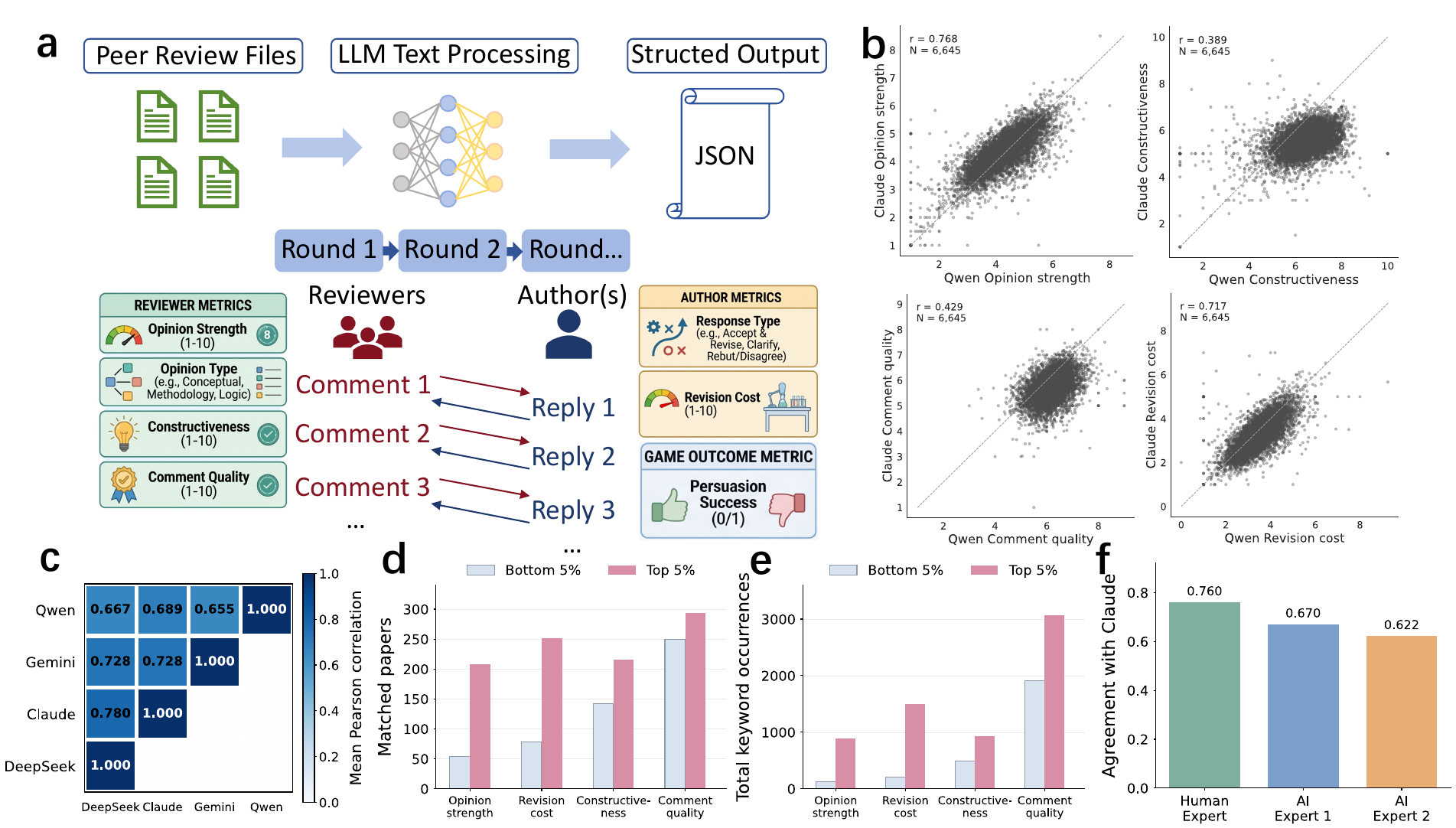}
  \caption{Validation of the AI-derived peer-review metrics. (a) Schematic overview of the pipeline that converts raw peer review correspondence into LLM-extracted comment--response pairs and structured review indicators. (b) Scatterplot comparisons between Claude and Qwen for opinion strength, constructiveness, comment quality, and revision cost; all correlations are significant. (c) Mean pairwise cross-model correlations across Claude Sonnet 4.6, Qwen-3.5-27B, Gemini-3-Flash, and DeepSeek-V3.2-nothinking for seven extracted variables, including the four continuous metrics as well as total rounds, total reviewers, and comment count. (d,e) Keyword-based validation comparing top- and bottom-tail papers on each continuous metric. (f) Agreement between Claude-derived rankings and three expert workflows on first-round author-level summaries.}
  \label{fig:fig1}
\end{figure}

Peer review correspondence is long and unstructured, so we begin by converting it into discrete reviewer concerns and aligned author replies (Fig.~\ref{fig:fig1}a). For each paper, we feed the full correspondence---all reviewer comments and author responses across all available rounds---into a fixed-prompt extraction pipeline. We then decompose reviewer text into independently addressable comment points, align the corresponding author responses, assign separate scores for opinion strength, constructiveness, comment quality, and revision cost, and track whether the same issue returns in later rounds. This procedure also produces paper-level quantities such as the number of review rounds, the number of reviewers (Fig.~\ref{fig:fig2}a, Supplymentary Fig.1), and the total number of extracted comment points. We first ask whether these measurements are reproducible across models. The four continuous metrics are positively correlated between Claude and Qwen (Fig.~\ref{fig:fig1}b), and when we repeat the extraction with four model families---Claude Sonnet 4.6, Qwen-3.5-27B, Gemini-3-Flash, and DeepSeek-V3.2-nothinking---the average pairwise correlations remain positive and significant for all seven extracted quantities (i.e., the pearson correlations between each pair are all above 0.65. Fig.~\ref{fig:fig1}c). These cross-model checks suggest that the pipeline is not simply reproducing idiosyncratic outputs from a single model.

We next ask whether the extracted metrics are interpretable in the underlying text. To do this, we validate the four continuous measures in two complementary ways. First, we use metric-specific keyword sets generated with both human and AI input (Supplementary Table 3) and compare the upper and lower tails of each distribution. The expected keywords appear more often in the upper tails than in the lower tails for all four dimensions (Fig.~\ref{fig:fig1}d,e), indicating that high-scoring cases are associated with the language the metrics are meant to capture. Second, we compare Claude-derived outputs with three expert workflows on first-round author-level summaries (Fig.~\ref{fig:fig1}f). One workflow is fully manual, and two are AI-assisted, but all three follow the same evaluation logic. In this validation, each workflow identifies which sampled papers fall at the top and bottom of the four continuous metrics, and we convert Claude's outputs into the same high--low ranking task before measuring agreement. We find that Claude-derived rankings align well with the rankings produced by these expert workflows. Taken together, these checks indicate that the pipeline captures stable and interpretable features of the correspondence and provide the basis for the downstream analyses.

\subsection{Review pressure is concentrated in the first round and on core claims}

\begin{figure}[p]
  \centering
  \includegraphics[width=\textwidth]{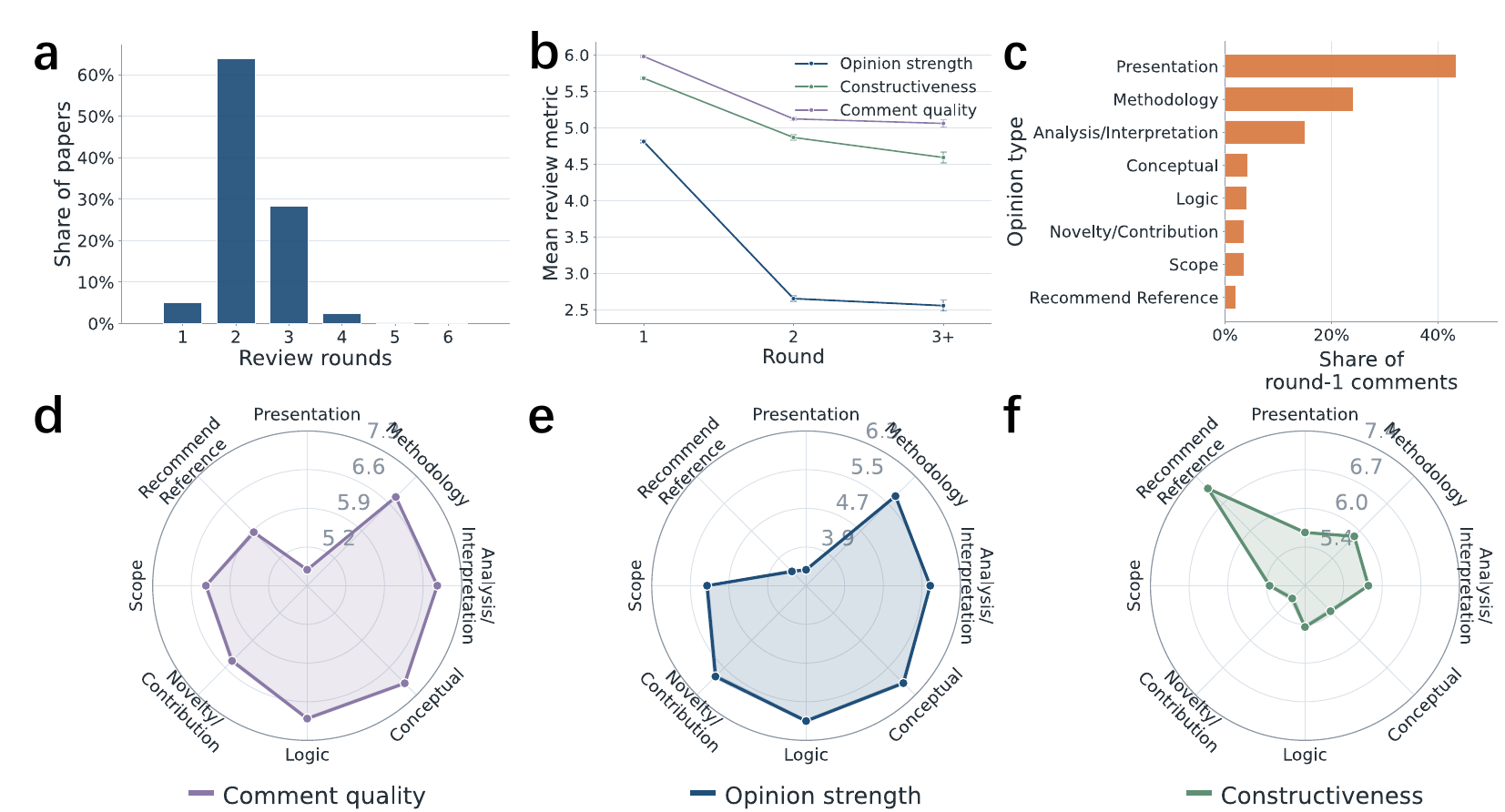}
  \caption{Baseline structure of review rounds and opinion types. (a) Distribution of the number of review rounds per paper. (b) Round-level changes in opinion strength, constructiveness, and comment quality. (c) Composition of first-round comments by opinion type. (d--f) Differences across opinion types in opinion strength, constructiveness, and comment quality. The figure highlights that frequency is not consequence: presentation comments are most common, but the strongest and highest-quality comments concentrate on conceptual, logical, and novelty-related issues.}
  \label{fig:fig2}
\end{figure}

\begin{figure}[p]
  \centering
  \includegraphics[width=\textwidth]{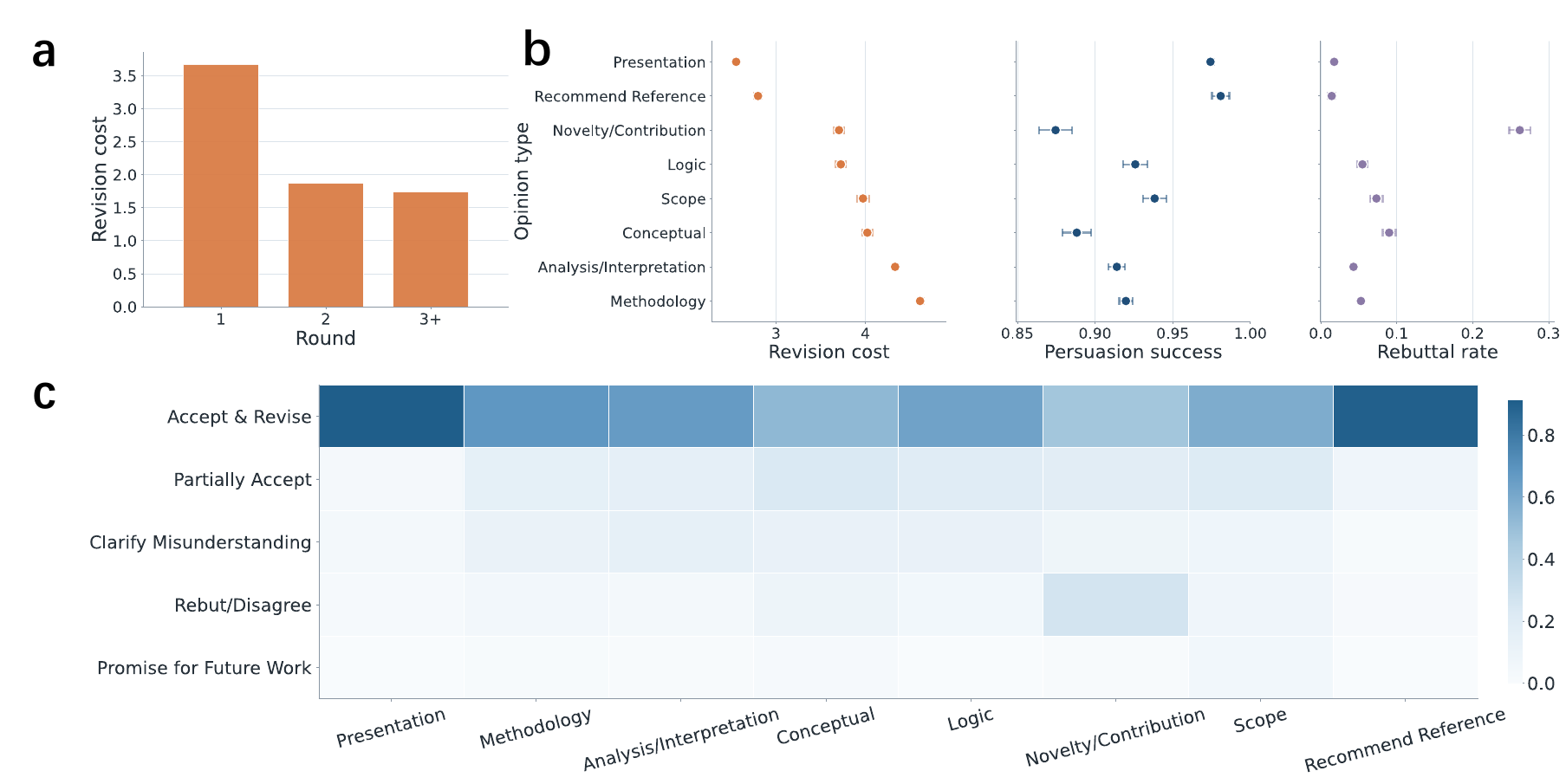}
  \caption{Author responses and revision burden. (a) Mean revision cost by review round. (b) Revision cost, persuasion success, and rebuttal rate across opinion types. (c) Distribution of author response types within each opinion type. The figure shows that revision burden is highest for methodology and analysis comments, whereas negotiation concentrates on core-claim criticisms, especially novelty/contribution, rather than on peripheral requests.}
  \label{fig:fig3}
\end{figure}

It is natural to expect the first review round to carry the greatest pressure, and our data quantitatively confirms this pattern. We average issue-level scores within rounds and classify each extracted comment by the part of the paper it targets---concepts, methodology, analysis/interpretation, logic, novelty/contribution, scope, presentation, or recommended references. We classify author responses in parallel by whether authors accept and revise, clarify misunderstanding, rebut/disagree, partially accept, or promise future work, and we use whether an issue reappears in later rounds as an operational proxy for persuasion success. We also define rebuttal rate as the share of responses classified as rebut/disagree. We begin with the overall round structure. Most papers go through two or three rounds, with two rounds accounting for about 64\% of the sample and three rounds accounting for about 28\% (Fig.~\ref{fig:fig2}a). Yet the pressure of review is concentrated much earlier. Opinion strength, constructiveness, and comment quality all fall sharply after round 1 (Fig.~\ref{fig:fig2}b), and revision cost follows the same pattern, dropping from its highest level in the first round to much lower levels thereafter (Fig.~\ref{fig:fig3}a). We therefore see that later rounds are usually not where the main substantive confrontation happens. They are more often the stage at which earlier disputes are narrowed, clarified, or wrapped up.

We next ask what kinds of issues dominate this early exchange. Presentation comments are the most common in round 1, followed by methodology and analysis/interpretation comments (Fig.~\ref{fig:fig2}c). But frequency and consequence are not the same thing. The strongest comments are concentrated in conceptual, logic, and novelty/contribution critiques, with methodology and analysis/interpretation also elevated relative to presentation (Fig.~\ref{fig:fig2}d). Comment quality follows a similar ordering, with the highest values concentrated in comments that challenge central concepts, logic, and interpretation rather than surface form (Fig.~\ref{fig:fig2}f). By contrast, constructiveness is highest for recommend-reference comments and also relatively high for methodology and analysis/interpretation comments, while novelty/contribution comments are among the least constructive (Fig.~\ref{fig:fig2}e). In other words, the comments that are easiest to act on are not necessarily the comments that are most consequential for the paper's core claims.

We then examine how authors respond to these different kinds of criticism. Here again, the most visible comments are not the ones that generate the hardest exchange. Revision cost is highest for methodology and analysis/interpretation comments and lowest for presentation comments, suggesting that the most labor-intensive revisions are tied to evidence, analysis, and research design rather than to wording alone (Fig.~\ref{fig:fig3}b). Persuasion success is highest for recommend-reference and presentation comments, whereas rebuttal rates peak for novelty/contribution comments and remain elevated for conceptual and scope-related issues (Fig.~\ref{fig:fig3}b). The distribution of author response types shows the same pattern: presentation and recommend-reference comments are usually handled through accept-and-revise, whereas comments on contribution, concepts, and scope more often generate clarification, partial acceptance, or rebuttal (Fig.~\ref{fig:fig3}c). We therefore find that the early rounds of peer review are not dominated simply by more comments. They are dominated by comments that more directly test what the paper claims to contribute and how far those claims can be sustained.

\subsection{Reviewer Disagreement and Review Fairness}

\begin{figure}[p]
  \centering
  \includegraphics[width=\textwidth]{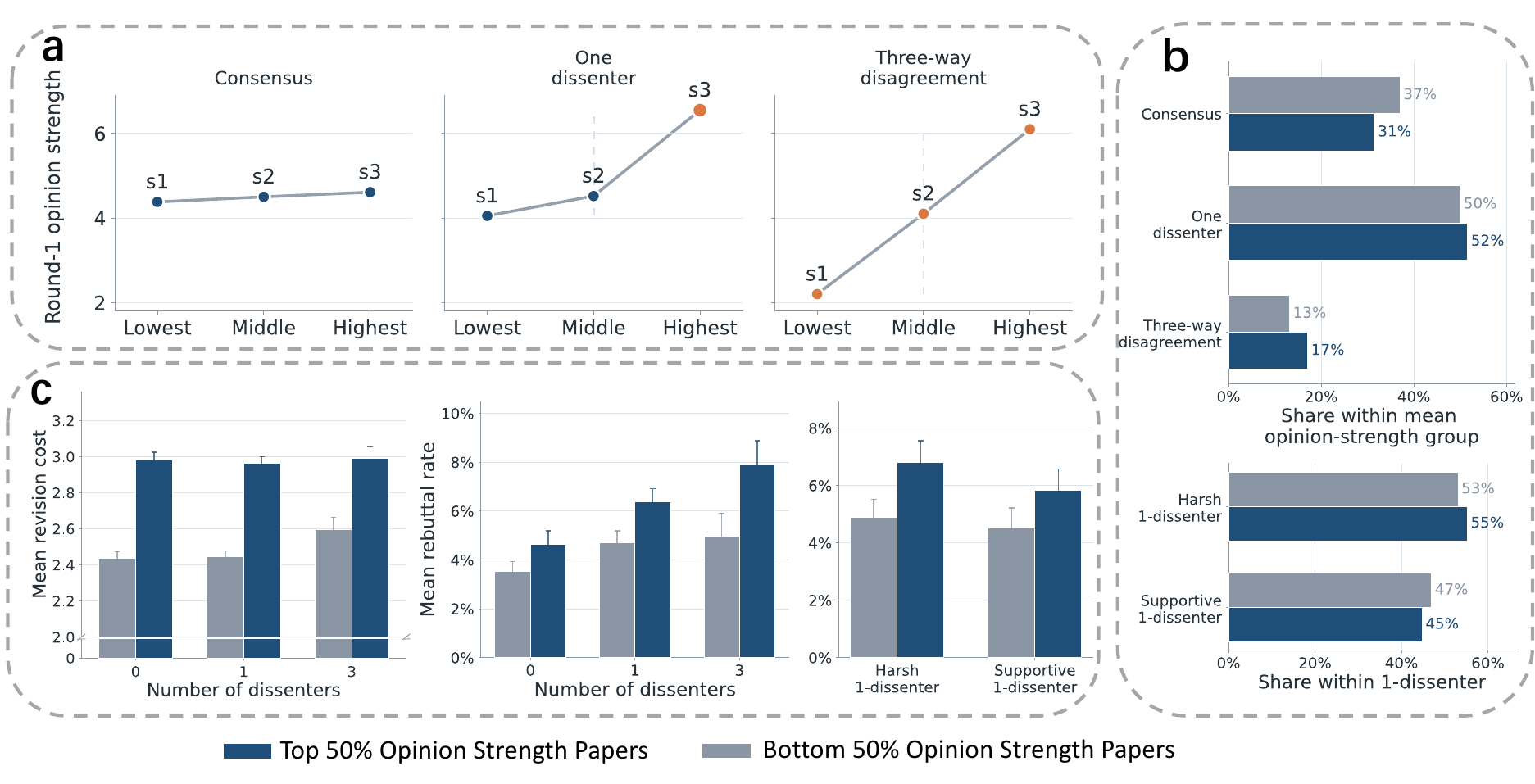}
  \caption{Reviewer-consensus configurations among papers with exactly three first-round reviewers. For each paper, we compare the three reviewers' mean first-round opinion-strength scores using a consensus threshold of 1. (a) Schematic examples of full consensus, one dissenter, and no consensus. (b) Distribution of these three configurations in lower- and higher-opinion-strength groups (top), together with the supportive-versus-harsh split within the one-dissenter cases (bottom). As average opinion strength rises, the share of papers with a dissenter increases. The lower panel further shows how one-dissenter cases are distributed between supportive and harsh outliers across the two groups. (c) Left: mean revision cost by number of dissenters in lower- and higher-opinion-strength groups. Right: rebuttal probability by number of dissenters, with the one-dissenter cases further separated into supportive and harsh dissenters. Stronger review is associated with more internal reviewer disagreement, while rebuttal becomes more likely as disagreement increases.}
  \label{fig:fig4}
\end{figure}

Strong review does not always arrive as a unified signal. Some papers receive three closely aligned evaluations, whereas others face a clear split among reviewers. To distinguish these patterns, we restrict the analysis to papers with exactly three first-round reviewers. For each paper, we calculate each reviewer’s mean first-round opinion strength and compare the three pairwise differences. Using a consensus threshold of 1, we classify the paper as full consensus when all reviewer pairs fall within that threshold, as one dissenter when one reviewer is separated from the other two, and as no consensus when no stable pair remains within the threshold. Figure~\ref{fig:fig4}a illustrates these three configurations schematically. We then divide papers into lower- and higher-opinion-strength groups and compare how these configurations are distributed. The overall distribution shifts toward disagreement as average opinion strength rises: dissenter configurations account for a larger share of papers in the higher-opinion-strength group (Fig.~\ref{fig:fig4}b, top). Within the one-dissenter cases, we further distinguish whether the outlying reviewer is relatively supportive or relatively harsh. Figure~\ref{fig:fig4}b also shows how that one-dissenter subset is distributed across the two groups. We therefore find that stronger review is not only more severe on average; it is also more likely to contain disagreement across reviewers.

These disagreement structures are also reflected in how authors respond. When average opinion strength is already high, revision cost remains high regardless of whether there are zero, one, or two dissenters. When average opinion strength is lower, however, revision cost rises with the number of dissenters (Fig.~\ref{fig:fig4}c, left). Rebuttal probability shows an even clearer pattern: in both lower- and higher-opinion-strength groups, the probability of rebuttal increases as the number of dissenters increases (Fig.~\ref{fig:fig4}c, right). Moreover, within the one-dissenter cases, rebuttal is more likely when that single dissenter is harsh rather than supportive. These results suggest that disagreement matters not only because it marks a lack of reviewer convergence, but also because it changes how authors respond. Mixed reviews with a clearly harsh outlying reviewer are more likely to trigger explicit author pushback than reviews in which the outlier is comparatively supportive.

A related question is whether broader author-team characteristics are similarly associated with the substance of review. Table~\ref{tab:figure4_correlations} reports paper-level Spearman correlations between team size, institution count, average career age decile, maximum career age decile, and the extracted review and response measures, visualized in Supplementary Fig.~2. Overall, broad team attributes show little association with the substance of review. Most of these associations are small, especially for opinion strength, comment quality, and persuasion success. The clearest exception is revision cost. It is higher for larger teams and lower for teams with older average or maximum career age, with institution count showing only a modest positive association. We therefore do not find evidence that coarse team attributes strongly structure the substantive content of review. Instead, the main detectable team-level pattern is that larger and younger teams appear to absorb more revision work once review is underway. Broad author-team characteristics do little to predict the substance of review, which is consistent with a relatively impartial review process in which reviewers respond more to the manuscript than to coarse features of the author team.

\subsection{Higher-impact papers receive stronger review}

\begin{figure}[p]
  \centering
  \includegraphics[width=\textwidth]{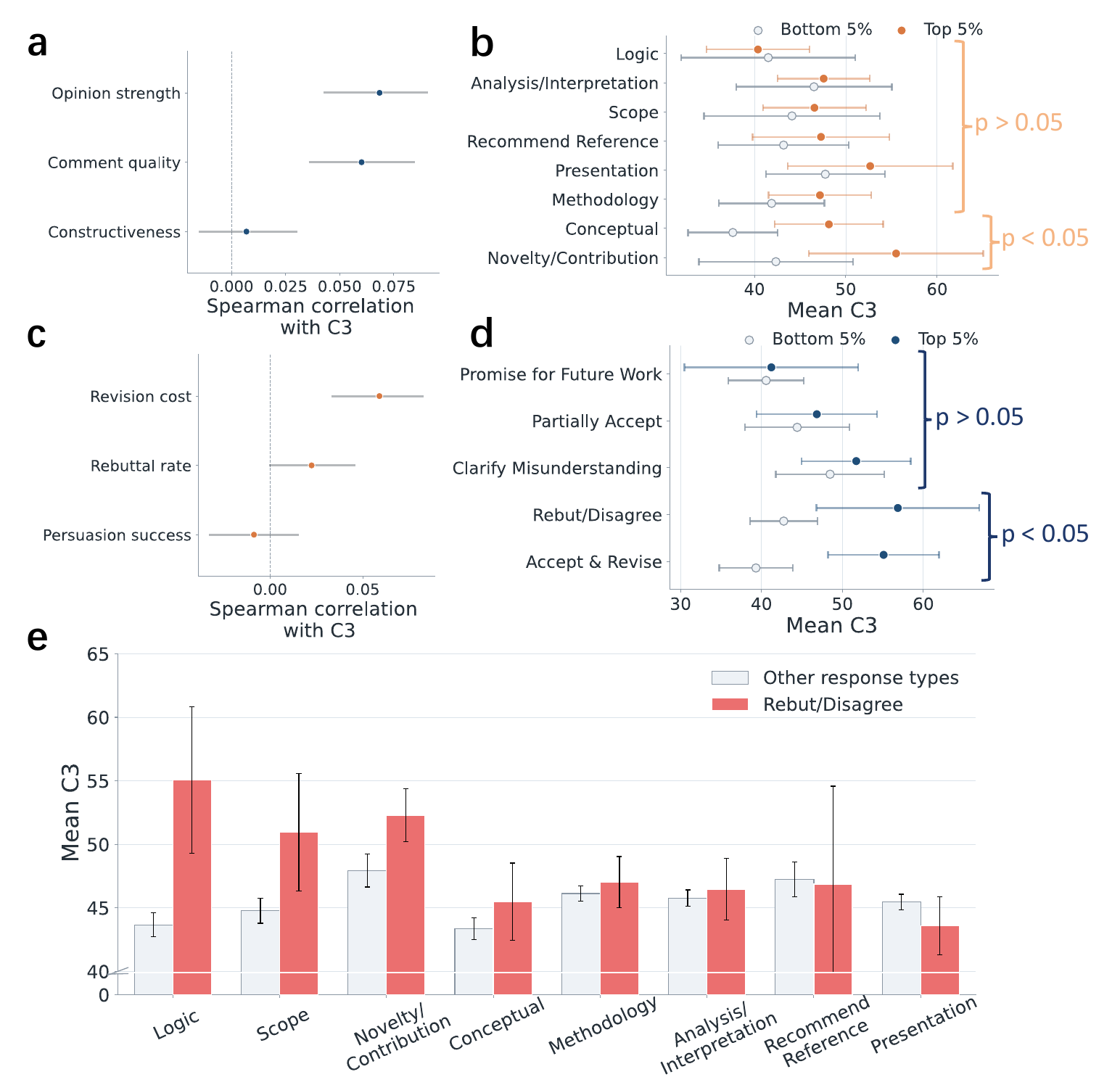}
  \caption{Review characteristics and later citation impact. (a) Spearman correlations between review metrics and the paper-level impact indicator $C_3$. (b) Differences in mean $C_3$ between papers in the top and bottom 5\% of opinion strength within year $\times$ opinion-type strata. (c) Correlations between response metrics and $C_3$. (d) Differences in mean $C_3$ between papers in the top and bottom 5\% of revision cost within year $\times$ response-type strata. (e) Mean $C_3$ for rebut/disagree versus other response types within each opinion type. Within accepted papers, stronger and higher-quality review, together with higher revision cost, are associated with higher later citation impact, whereas the issue-resolution proxy remains close to null.}
  \label{fig:fig5}
\end{figure}

A natural question is whether, within accepted papers, those with greater later impact pass through review more easily. We address this by aggregating issue-level measures to the paper level and relating these paper-level summaries to $C_3$, the citation count within three years after publication, which provides a practical early-impact window: long enough for meaningful citation differences to emerge, but short enough to remain usable for recent publication years. We begin with the overall correlations in Fig.~\ref{fig:fig5}. The pattern already runs against the smooth-passage intuition. Opinion strength and comment quality are positively associated with $C_3$, whereas constructiveness remains close to zero (Fig.~\ref{fig:fig5}a). On the response side, revision cost is also positively associated with $C_3$, while persuasion success and rebuttal rate remain close to null (Fig.~\ref{fig:fig5}c). We then check whether the same pattern appears in simpler profile comparisons. In Supplementary Fig.~3, papers in the top 10\% of the $C_3$ distribution score higher on opinion strength, comment quality, and revision cost than papers in the bottom 10\%, whereas constructiveness is only slightly lower and the issue-resolution proxy and rebuttal rate remain similar. We next ask whether the main association is driven by one publication year. Supplementary Fig.~4 suggests that revision cost is positive in most years and especially strong in 2021, comment quality is positive in nearly every year, and opinion strength is positive in most years except 2017, whereas constructiveness, persuasion success, and rebuttal rate remain weak or unstable.

We further examine whether these associations survive after adjusting for other paper-level differences. In the Ordinary Least Squares regression (OLS) models (Supplementary Table 5-9), we regress $\log(1 + C_3)$ on one focal predictor at a time while controlling for team size, institution count, team average and maximum career age, total review rounds, total reviewers, and year fixed effects. The same profile persists: opinion strength, comment quality, and revision cost remain positively associated with later impact, whereas constructiveness, persuasion success, and rebuttal rate remain weak or close to null. Among the overall OLS coefficients, revision cost is the strongest positive term. Taken together, these checks show that higher-impact papers are not the papers that move through review with the least friction. Rather, within accepted papers, they are more often the papers that attract stronger criticism, more intellectually substantial comments, and more extensive revision.

The review--impact association also varies across kinds of scrutiny. To answer this, we compare papers in the top and bottom 5\% of opinion strength within the same year $\times$ opinion-type stratum and papers in the top and bottom 5\% of revision cost within the same year $\times$ response-type stratum. The resulting heterogeneity is substantial. The largest gaps in $C_3$ appear when strong criticism concerns novelty/contribution and conceptual issues, whereas some other categories, such as logic, show little or no advantage in the main comparison (Fig.~\ref{fig:fig5}b). The supplementary regressions point in the same direction. Positive coefficients for opinion strength are concentrated in novelty/contribution, analysis/interpretation, and conceptual comments, while presentation, scope, logic, recommended references, and methodology do not show similarly strong positive terms in the type-specific models. We therefore find that strong review is most informative when it is aimed at the claims that define what the paper is trying to add.

We further examine the response side of the same question. The separation by revision cost is clearest for accept-and-revise and rebut/disagree profiles (Fig.~\ref{fig:fig5}d), and the supplementary regression results again show positive coefficients for these two response types but not for partial acceptance or future-work promises. Rebuttal is also informative only in some contexts, with rebut/disagree responses associated with higher $C_3$ within logic, scope, novelty/contribution, and conceptual comments, but not within presentation or recommend-reference comments (Fig.~\ref{fig:fig5}e). This pattern suggests that disagreement is not simply a generic sign of a difficult review. It becomes informative when reviewers and authors are contesting the paper's core contribution, conceptual framing, logical structure, or interpretive scope. Taken together, these results suggest that the review--impact link is strongest when scrutiny and disagreement center on consequential claims.

\subsection{Peer review differs more across fields in style than in length}

\begin{figure}[p]
  \centering
  \includegraphics[width=\textwidth]{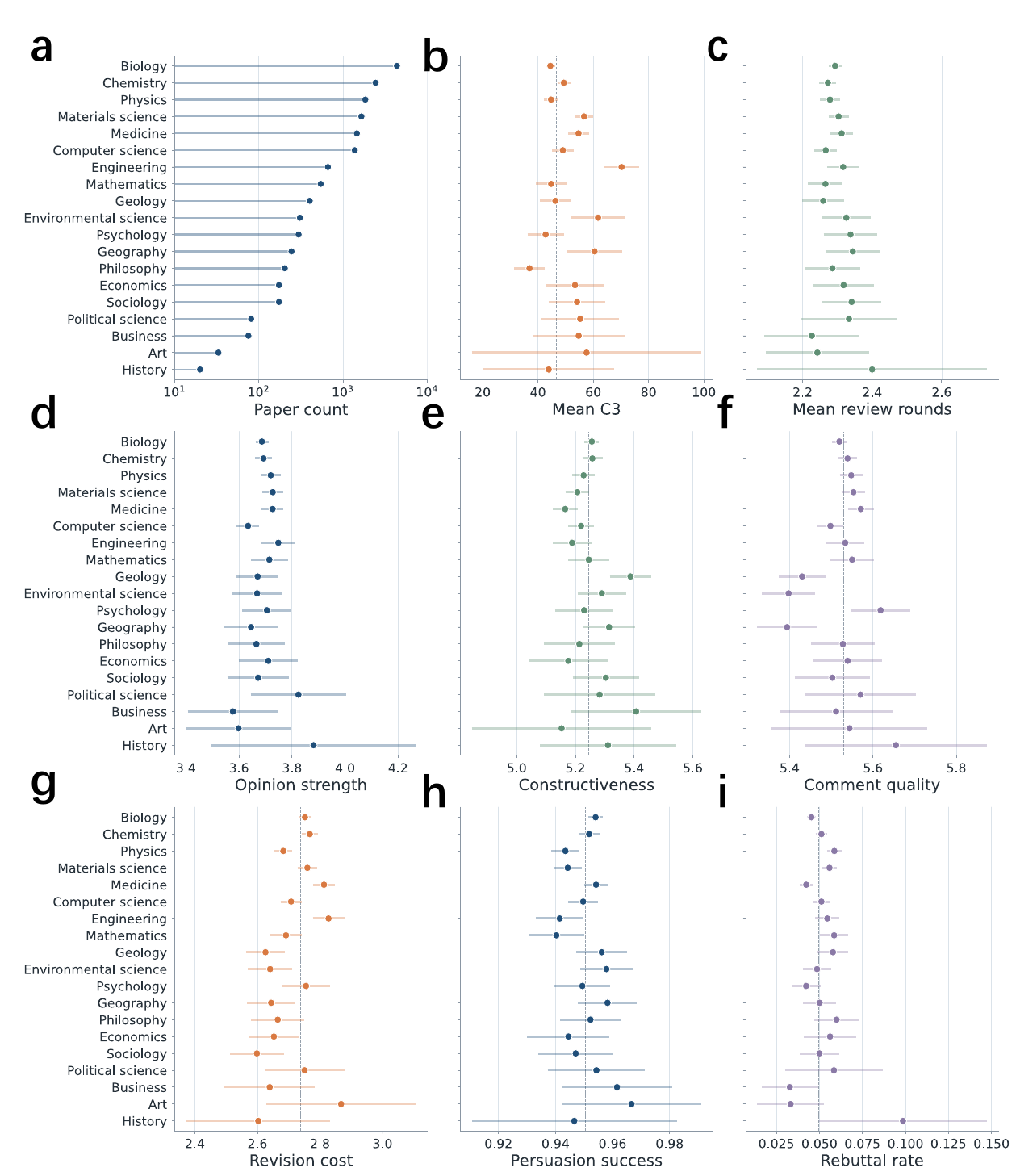}
  \caption{Disciplinary heterogeneity in peer review and paper impact. Panels compare paper counts, mean $C_3$, mean review rounds, opinion strength, constructiveness, comment quality, revision cost, the issue-resolution proxy, and rebuttal rate across fields. While average review-round counts vary relatively little, the substantive profile of peer review differs markedly across disciplines; field-level contrasts should therefore be interpreted together with unequal sample sizes across disciplines.}
  \label{fig:fig6}
\end{figure}

We next turn directly to disciplinary context. In Fig.~\ref{fig:fig6}, average review-round counts cluster tightly, at roughly 2.2 to 2.4 rounds across fields, so review length itself varies little. Review style, however, varies much more. Opinion strength is higher in fields such as history and political science than in business or art. Revision cost is higher in engineering, medicine, and art than in sociology or geology. Rebuttal is more common in history, philosophy, and physics than in business, art, or medicine. Constructiveness and persuasion success also vary across fields. Constructiveness is relatively high in business, geology, and geography but lower in art, medicine, and economics, while persuasion success is relatively high in art, business, geography, and environmental science but lower in mathematics, engineering, and physics. Taken together, these results suggest that peer review varies less with who the authors are than with the disciplinary context in which evaluation takes place. In other words, fields differ more clearly in their styles of criticism, revision, persuasion, and rebuttal than in the overall length of review.

\section{Discussion}

In this paper, we find that within the accepted-paper sample, demanding review is associated with higher later impact rather than with papers moving smoothly through evaluation. Papers with higher $C_3$ receive stronger criticism, higher-quality comments, and greater revision cost, while constructiveness, persuasion success, and rebuttal rate remain much less informative. These associations are strongest when scrutiny targets consequential issues---especially novelty/contribution, concepts, interpretation, logic, and scope---rather than peripheral presentation. Among published papers, intense review therefore appears to mark concentrated scrutiny of ambitious or consequential claims, not only manuscript weakness.

This changes how difficult review should be interpreted by authors. Receiving strong criticism is not necessarily bad news. In our data, the papers that later become more influential are often those that attract sharper challenges and require more substantial revision. The response that appears most informative is not blanket compliance or blanket resistance, but a response matched to the type of criticism. When reviewers point to presentation problems, missing references, or localized clarifications, straightforward revision is usually enough. When reviewers challenge the paper's core contribution, conceptual framing, logic, or interpretive scope, however, authors may need to respond more substantively by narrowing overstatement, adding new evidence, clarifying the intended claim, or defending a central argument explicitly and with support. The results are especially suggestive here. Rebuttal is not broadly associated with higher impact, but it is more strongly associated with higher impact when disagreement concerns consequential claims rather than peripheral ones. A clear, evidence-backed defense of a central claim is therefore not necessarily a warning sign; within accepted papers, it can be part of the review profile associated with higher-impact work.

The analysis also clarifies what kind of scrutiny matters most and where it varies. Presentation comments are the most common, yet the exchanges that appear most consequential concern core claims and are more likely to generate clarification, partial acceptance, or rebuttal rather than simple compliance. Among papers reviewed by three first-round reviewers, stronger review is also more likely to contain dissenter configurations, and rebuttal becomes more common as the number of dissenters increases, especially when the lone dissenter is comparatively harsh. Broad team traits still explain little of the variation in review, which is consistent with a relatively fair process in which reviewers respond more to the manuscript than to broad features of the author team. Field differences are much more pronounced. Average review length changes little across disciplines, but review style does not. Criticism is stronger in history and political science than in business or art, revision cost is higher in engineering, medicine, and art than in sociology or geology, and rebuttal is more common in history, philosophy, and physics than in business, art, or medicine. Constructiveness is also relatively high in business, geology, and geography, whereas persuasion success is higher in art, business, geography, and environmental science. Taken together, these patterns suggest that peer review is relatively impartial to broad team traits but remains shaped by disciplinary styles of criticism, revision, persuasion, and rebuttal.

These findings have implications for journals and peer-review policy. Journals may benefit from directing reviewer attention more explicitly toward the manuscript's central claims, especially contribution, conceptual framing, and interpretation. Editorial guidelines and review forms could be designed to reinforce this priority. Journals should also be cautious about treating smooth or low-friction review as a sufficient indicator of review quality; a more useful standard is whether peer review engages substantively with the claims that matter most. Because review practices vary across disciplines, reforms and evaluation criteria should avoid one-size-fits-all benchmarks.

These conclusions should be interpreted within the design of the study. The data include only papers that were ultimately published, so the analysis cannot identify how review differs between accepted and rejected submissions or whether stricter review causes higher impact. The issue-resolution measure is also an operational proxy based on cross-round persistence rather than a direct measure of reviewer intent. Within those limits, the study shows that open peer review correspondence can be converted into structured issue-level data and used to study how criticism, response, and revision unfold before publication. This creates a basis for more process-oriented research on peer review, including comparisons across journals, rejected-paper trajectories, reviewer-level heterogeneity, and the allocation of scrutiny across different kinds of scientific claims.

\section{Methods}

\subsection{Data}

The analysis was based on open peer review correspondence associated with research articles published in \textit{Nature Communications} between 2017 and 2024. For each year, 1,000 papers were randomly sampled, yielding a year-balanced corpus of 8,000 papers. For each paper, the dataset included the complete reviewer comments across all available review rounds and the corresponding author response letters. The corpus therefore captured the full revision trajectory of accepted papers rather than a single editorial snapshot, making it possible to interpret review intensity conditional on eventual publication.

To relate review dynamics to later scholarly impact, paper-level citation impact was measured using the $C_3$ indicator from SciSciNet-V2 \cite{lin2023sciscinet}. Because the citation observation window is shorter for recent papers, the 2022--2024 cohort can only be observed up to the time covered by that database. This limitation was taken into account when interpreting impact-related analyses.

In addition to the review correspondence, the paper-level analyses used team-related variables, including team size, institution count, team average career age decile, and team maximum career age decile, as well as disciplinary groupings for field-level comparisons.

\subsection{LLM-based extraction of structured review data}

For each paper, we gave the large language model the complete peer review file, including all reviewer comments and all author responses from every available review round, and asked it to process the material using a fixed prompt. The model was instructed to return a single structured JSON file containing one object for each reviewer in each round. Within each reviewer-round object, the model was asked to segment the reviewer text into discrete, independently addressable comment points, identify the author response addressing each point when possible, and assign predefined scores and labels to each extracted comment--response pair.

This procedure yielded an LLM-derived representation of reviewer--author interaction at the issue level. We treat these extracted units as operationalized measurements rather than ground-truth annotations. In particular, comment decomposition, response matching, and cross-round tracking were all inferred by the model from the correspondence text under fixed prompt rules rather than produced by deterministic parsing or manual coding.

\subsection{Extracted variables}

Each extracted comment point was assigned four ordinal scores ranging from 1 to 10: opinion strength, constructiveness, comment quality, and revision cost. Opinion strength captured the severity of the problem identified by the reviewer; constructiveness captured the degree to which the comment provided actionable guidance; comment quality captured the intellectual and scholarly value of the comment; and revision cost captured the amount of work the author reported undertaking or promising in response. Detailed guiding questions and score anchors for these four continuous metrics are reported in the Supplementary Information Table 2,3.

Each comment point was also assigned two single-label categorical variables: opinion type and author response type. Opinion type classified the main focus of the reviewer comment, including conceptual issues, methodology, analysis/interpretation, logic, novelty/contribution, scope, presentation, direct acceptance, and recommended references. Direct-acceptance cases were defined during extraction but were not emphasized in the main comparative figures, which focus on substantive criticism categories. Author response type classified the author's primary response strategy, including accept-and-revise, clarify misunderstanding, rebut/disagree, partially accept, and promise for future work. The full classification priorities and disambiguation rules are reported in the Supplementary Information B.

Throughout the paper, \textbf{demanding review} is used as a descriptive umbrella term for combinations of stronger criticism, higher comment quality, and greater revision cost; it is not a separate extracted variable. The scoring scheme was explicitly decoupled across dimensions. In particular, severity, constructiveness, quality, and revision burden were evaluated as distinct attributes rather than collapsed into a single holistic judgment.

Let $p$ index LLM-extracted comment--response pairs. For each pair, the extracted representation can be written as
\begin{equation}
X_p = \left(s_p, c_p, q_p, r_p, o_p, a_p, u_p\right),
\end{equation}
where $s_p$ is opinion strength, $c_p$ is constructiveness, $q_p$ is comment quality, $r_p$ is revision cost, $o_p$ is opinion type, $a_p$ is author response type, and $u_p \in \{0,1\}$ is the persuasion-success indicator inferred from whether the issue reappears in later rounds.

\subsection{Aggregation and statistical analysis}

The primary unit of extraction was the issue-level comment point. For paper-level analyses, continuous variables were aggregated by taking the mean across extracted comment points within each paper. Round-level and field-level summaries were obtained by averaging within the corresponding subsets, and categorical variables were summarized as proportions.

Associations between extracted variables and paper-level outcomes were evaluated primarily using Spearman rank correlations. In addition, extreme-group comparisons contrasted papers in the upper and lower tails of the relevant distributions within stratified comparison sets such as year $\times$ opinion type or year $\times$ author response type. All analyses are descriptive or correlational.

\subsection{Validation}

The extracted metrics were validated using three complementary strategies.

First, cross-model consistency was examined by comparing outputs from multiple language models. Pairwise correlations were computed across models for the four continuous metrics and several aggregate quantities, including opinion strength, revision cost, constructiveness, comment quality, total rounds, total reviewers, and comment count. These comparisons assessed whether the extracted patterns were robust to model choice.

Second, keyword-based validation was conducted for the four continuous metrics. Human experts and AI systems generated metric-relevant keyword sets corresponding to high values of each dimension. The prevalence of these keywords was then compared between papers in the top and bottom tails of the metric distributions.

Third, expert-based validation was performed on author-level summaries from the first review round. All three expert workflows were designed around the same evaluation logic. The human expert followed this logic directly as a manual reading protocol. AI expert 1 used the same logic to generate and execute a stepwise workflow, whereas AI expert 2 used a one-shot prompt based on the same logic to produce comparable outputs. For each workflow, the resulting labels were compared with Claude-derived outputs after the Claude extractions had been transformed into the same comparable format, and agreement was quantified as accuracy. Together, these validation steps establish reproducibility and interpretability of the extracted indicators, rather than a ground-truth verification of every decomposition and alignment decision.

\section*{Acknowledgments}
This work is supported by the National Natural Science Foundation of China under grant T2541030 and 72274020, and the Fundamental Research Funds for the Central Universities, grant 2233200016.

\section*{Data availability}
This study uses publicly available peer review files published by \textit{Nature Communications}. Citation-impact data were drawn from the $C_3$ measure in SciSciNet-V2 \cite{lin2023sciscinet}, which is available at \url{https://northwestern-cssi.github.io/sciscinet/}.

\section*{Author contributions}
HJ and AZ designed the research. HJ performed the experiments. HJ and AZ analyzed the data. HL and ZW provided analytical tools and methods. All authors wrote the manuscript.

\bibliographystyle{unsrt}
\bibliography{references}

\clearpage

\clearpage

\clearpage

\clearpage

\clearpage
\begin{table}[p]
\centering
\small
\caption{Spearman correlations between team characteristics and extracted review/response metrics}
\label{tab:figure4_correlations}
{\renewcommand{\arraystretch}{1.35}
\begin{tabular*}{0.96\textwidth}{@{\extracolsep{\fill}}lccccc@{}}
\toprule
\rule{0pt}{2.8ex}\shortstack[c]{Explanatory\\variable} & \shortstack[c]{Opinion\\strength} & \shortstack[c]{Constructive-\\ness} & \shortstack[c]{Comment\\quality} & \shortstack[c]{Revision\\cost} & \shortstack[c]{Persuasion\\success} \\[4pt]
\midrule
\shortstack[l]{Team size\\\phantom{***}} & \shortstack[c]{0.04\\[1pt]***} & \shortstack[c]{-0.04\\[1pt]**} & \shortstack[c]{0.05\\[1pt]***} & \shortstack[c]{0.11\\[1pt]***} & \shortstack[c]{-0.01\\[1pt]\phantom{***}} \\
\addlinespace[2pt]
\shortstack[l]{Institution count\\\phantom{***}} & \shortstack[c]{0.02\\[1pt]\phantom{***}} & \shortstack[c]{-0.02\\[1pt]\phantom{***}} & \shortstack[c]{0.02\\[1pt]\phantom{***}} & \shortstack[c]{0.04\\[1pt]**} & \shortstack[c]{0.00\\[1pt]\phantom{***}} \\
\addlinespace[2pt]
\shortstack[l]{Average career\\age decile} & \shortstack[c]{-0.01\\[1pt]\phantom{***}} & \shortstack[c]{0.00\\[1pt]\phantom{***}} & \shortstack[c]{0.00\\[1pt]\phantom{***}} & \shortstack[c]{-0.07\\[1pt]***} & \shortstack[c]{0.00\\[1pt]\phantom{***}} \\
\addlinespace[2pt]
\shortstack[l]{Maximum career\\age decile} & \shortstack[c]{0.01\\[1pt]\phantom{***}} & \shortstack[c]{-0.01\\[1pt]\phantom{***}} & \shortstack[c]{0.00\\[1pt]\phantom{***}} & \shortstack[c]{-0.06\\[1pt]***} & \shortstack[c]{0.00\\[1pt]\phantom{***}} \\
\bottomrule
\end{tabular*}}

\vspace{2pt}
\raggedright\footnotesize Significance levels: * $p < 0.05$, ** $p < 0.01$, *** $p < 0.001$.
\end{table}

\clearpage

\clearpage

\clearpage
\setcounter{section}{0}
\setcounter{subsection}{0}
\setcounter{subsubsection}{0}
\setcounter{figure}{0}
\setcounter{table}{0}
\setcounter{equation}{0}

\begin{center}
{\Large\bfseries Supplementary Information: Demanding peer review is associated with higher impact in published science\par}
\vspace{0.5em}
{\normalsize \today\par}
\end{center}
\vspace{1em}

\section{Supplementary Information}

This Supplementary Information provides additional details on the prompt-based LLM extraction procedure, metric-extraction rubric, keyword-based validation, and supporting analyses for the main manuscript.

\subsection{Prompt-based extraction procedure}

For each paper, the full peer review correspondence---including all reviewer comments and author responses across all available review rounds---was provided to a large language model using a fixed prompt. The prompt instructed the model to return a single JSON array containing one object for each reviewer in each round. Within each reviewer-round object, the model was asked to segment the reviewer text into discrete comment points, identify the author response addressing each point when possible, and assign predefined scores and labels. The prompt also instructed the model to infer whether an issue was re-raised in later rounds.

These extracted units should be understood as LLM-derived operationalizations rather than deterministic parses or ground-truth annotations. The same prompt supplied the scoring anchors, decision rules, type-classification priorities, and required output schema.

To approximate whether a reviewer concern was resolved, we constructed a persuasion-success measure using a \emph{silence = acceptance} rule. If a reviewer explicitly repeated the same concern in a later round or expressed continued dissatisfaction, the issue was coded as unresolved. Otherwise, it was coded as resolved, including cases in which the concern was not mentioned again in subsequent rounds. Because this variable is inferred from prompted cross-round reading rather than directly observed reviewer intent, it should be interpreted as an operationalized persuasion-success measure rather than a literal measure of persuasion.

The four continuous metrics were extracted with a fixed 1--10 scoring rubric applied at the level of LLM-extracted comment--response pairs. The rubric was designed to separate conceptually distinct dimensions of reviewer--author interaction rather than collapse them into a single holistic judgment. In particular, \textit{opinion strength} and \textit{constructiveness} were scored independently: a severe criticism could still be highly constructive, and a minor point could be weakly actionable. Table~\ref{tab:si_metric_questions} summarizes the guiding question for each metric, and Table~\ref{tab:si_metric_anchors} reports the score anchors used in the extraction prompt.

\subsection{Categorical classification priorities and disambiguation}

In addition to the four continuous metrics, each extracted comment--response pair received one single-label opinion type and one single-label author response type. These labels were assigned by fixed prompt rules rather than post hoc interpretation.

\paragraph{Opinion type.} The prompt first checked two special cases: \textit{Accept}, used for pure endorsement without substantive criticism, and \textit{Recommend Reference}, used when the reviewer suggested specific paper(s) identifiable by author, year, or title. If neither special case applied, the prompt assigned the comment to the first matching substantive category in the following priority order: \textit{Conceptual}, \textit{Methodology}, \textit{Analysis/Interpretation}, \textit{Logic}, \textit{Novelty/Contribution}, \textit{Scope}, and \textit{Presentation}. These categories respectively captured challenges to the paper's premise or theoretical framing, data collection or design, statistical analysis or interpretation, inferential reasoning, claimed novelty relative to prior work, generalizability or boundary conditions, and clarity or presentation. The first-match rule was used to resolve overlap. For example, a request to cite a specific paper was classified as \textit{Recommend Reference}, whereas a generic claim that the literature review was incomplete without naming a paper was treated as \textit{Presentation}. Questions about how the study differed from prior work were classified as \textit{Novelty/Contribution}. Critiques of method choice affecting validity were classified as \textit{Methodology}, whereas critiques of statistical test choice were classified as \textit{Analysis/Interpretation}. Statements that the evidence did not support the conclusion were classified as \textit{Logic} rather than \textit{Analysis/Interpretation}.

\paragraph{Author response type.} The prompt assigned one primary response strategy to each author reply. \textit{Accept \& Revise} denoted full agreement accompanied by implemented or promised revision. \textit{Clarify Misunderstanding} denoted responses arguing that the reviewer had misread or overlooked material already in the manuscript. \textit{Rebut/Disagree} denoted explicit disagreement with the reviewer while maintaining the authors' position. \textit{Partially Accept} denoted mixed responses that accepted one part of the criticism but disputed another. \textit{Promise for Future Work} denoted acknowledgments that deferred the requested change beyond the scope of the current paper.

\paragraph{Persuasion success.} The prompt treated persuasion success as a binary, cross-round resolution indicator based on a \textit{silence = acceptance} rule. An issue was coded as unresolved (0) only when a later-round reviewer explicitly repeated the concern or expressed continued dissatisfaction, using signals such as ``this does not address my concern,'' ``the issue persists,'' or equivalent restatements of the same criticism. Otherwise the issue was coded as resolved (1), including cases with explicit acknowledgment (e.g., ``addressed'' or ``thank you for clarifying''), qualified acceptance, issue disappearance in later rounds, or cases with no subsequent review round. As in the main text, this variable should be interpreted as an operational proxy for issue resolution rather than a direct observation of reviewer intent.

\subsection{Aggregation of pair-level metrics}

The primary extraction operated at the level of LLM-extracted comment--response pairs, but several analyses required aggregation to higher levels. For any paper $i$ with comment--response pairs $\mathcal{P}_i$, the paper-level mean for a continuous metric $m_p$ was computed as
\begin{equation}
\bar{m}_i = \frac{1}{|\mathcal{P}_i|} \sum_{p \in \mathcal{P}_i} m_p.
\end{equation}
Round-level summaries were obtained by restricting $\mathcal{P}_i$ to pairs from the same review round, and field-level summaries were obtained by averaging across papers assigned to the same disciplinary group. For categorical variables, proportions within the relevant aggregation level were used. These aggregated measures were used to describe review-round dynamics, differences across opinion types, author-response profiles, disciplinary heterogeneity, and the relationship between team characteristics and review outcomes.

\subsection{Expert-workflow validation}

The expert-based validation used author-level summaries from the first review round and implemented the same underlying evaluation logic in three different ways. The human expert followed the protocol directly as a manual reading procedure. AI expert 1 used the same logic to generate a stepwise workflow and then executed that workflow. AI expert 2 used a one-shot prompt built from the same logic to generate the corresponding validation outputs in a single pass.

For each of the four continuous metrics, the expert workflows produced outputs in a common comparison format so that they could be aligned with the Claude-derived extraction results. The Claude-cleaned data were therefore converted into the same comparable format before evaluation. Agreement was then computed as accuracy between each expert workflow and the transformed Claude-derived outputs. This design allowed us to test whether the same validation logic yielded comparable judgments across manual execution, workflow-based AI execution, and one-shot AI execution.

\subsection{Keyword patterns for metric-aligned validation}

To supplement the model-consistency and expert-based validation analyses, we constructed metric-aligned keyword patterns for the four continuous dimensions: opinion strength, constructiveness, comment quality, and revision cost. These patterns were used to compare the textual signatures of papers located in the upper and lower tails of each metric distribution. The purpose of this exercise was not to recover a complete semantic representation of the metrics, but to test whether papers with higher extracted scores were also associated with interpretable language patterns that matched the intended meaning of each dimension.

Table~\ref{tab:si_keywords} summarizes the high-value keyword rules used in this analysis. These rules were used as supporting evidence for interpretability rather than as a replacement for the extracted metrics themselves.

\subsection{Team characteristics and extracted review metrics}

Supplementary Fig.~\ref{fig:sifig2} visualizes the relationship between team-level characteristics and the extracted peer-review indicators. The main result remains that most team-feature associations are small. The clearest exception concerns revision cost, which varies more systematically with team size and career-age composition than the other review and response metrics.

\subsection{Differences between high- and low-impact papers}

Supplementary Fig.~\ref{fig:sifig3} compares the review and response profiles of papers in the top 10\% and bottom 10\% of the $C_3$ distribution. High-$C_3$ papers score higher on opinion strength, comment quality, and revision cost, while constructiveness is slightly lower and the issue-resolution proxy and rebuttal-rate measures remain close to those of low-$C_3$ papers. This figure provides a compact summary of the broader impact-related findings presented in the main text.

\subsection{Year-by-year stability of the impact associations}

Supplementary Fig.~\ref{fig:sifig4} evaluates whether the relationship between review/response metrics and $C_3$ is stable across publication years. The figure shows that revision cost is positively associated with $C_3$ in most years and is especially strong in 2021. Comment quality is positive in nearly every year, and opinion strength is positive in most years except 2017. By contrast, constructiveness remains near zero or slightly negative, and the issue-resolution proxy and rebuttal-rate measures fluctuate around zero. These year-specific patterns support the interpretation that the main impact associations are not driven by a single year.

\subsection{OLS robustness analyses for the impact associations}

To complement the rank-correlation and extreme-group analyses in the main text, we estimated a set of supplementary ordinary least squares models using $\log(1 + C_3)$ as the dependent variable. Table~\ref{tab:overall_single_metric_ols} reports single-metric paper-level models, and Tables~\ref{tab:opinion_strength_by_type_ols}--\ref{tab:revision_cost_by_type_ols} report the corresponding type-specific models. In all OLS specifications, the control variables were team size, institution count, team average career age, team maximum career age, total review rounds, and total reviewers, together with year fixed effects implemented as $C(\mathrm{year})$. The generic specification can therefore be written as:
\begin{equation}
\begin{aligned}
\log(1 + C_3) \sim {}& \text{focal predictor} + \text{team size} + \text{institution count} + \text{team average career age} \\
& + \text{team maximum career age} + \text{total rounds} + \text{total reviewers} + C(\mathrm{year}).
\end{aligned}
\end{equation}
These supplementary regressions are intended as robustness checks rather than causal estimates. The overall pattern remains similar to the main analysis: opinion strength, comment quality, and revision cost show positive associations with later impact, whereas constructiveness, persuasion success (the issue-resolution proxy), and rebuttal rate remain weak or close to null.

\clearpage

\begin{table}[t]
\centering
\small
\caption{Guiding questions for the four continuous metrics}
\label{tab:si_metric_questions}
\begin{tabular}{@{}p{0.22\textwidth}p{0.28\textwidth}p{0.40\textwidth}@{}}
\toprule
Metric & Guiding question & Interpretation of higher scores \\
\midrule
Opinion strength & How severe is the identified problem? & The comment more directly threatens the validity, credibility, or acceptability of the manuscript. \\
Constructiveness & How actionable is the feedback? & The comment provides more concrete, specific, and implementable guidance for revision. \\
Comment quality & What is the intellectual merit of the comment? & The comment shows more specific, well-reasoned, and scientifically substantive engagement with the manuscript. \\
Revision cost & How much work did the author do or promise to do? & Addressing the comment requires more extensive rewriting, reanalysis, new evidence, or redesign. \\
\bottomrule
\end{tabular}
\end{table}

\begin{table}[p]
\centering
\small
\setlength{\tabcolsep}{4pt}
\caption{Score anchors used to extract the four continuous metrics}
\label{tab:si_metric_anchors}
\begin{tabular}{@{}p{0.06\textwidth}p{0.215\textwidth}p{0.215\textwidth}p{0.215\textwidth}p{0.215\textwidth}@{}}
\toprule
Score & Opinion strength & Constructiveness & Comment quality & Revision cost \\
\midrule
1 & Cosmetic issue, such as typography, spelling, or formatting. & Pure dismissal with no guidance for repair. & Factually wrong or based on a clear misreading. & Trivial correction, such as fixing a typo or spelling. \\
2 & Surface-level polish issue, such as awkward phrasing or word choice. & States a problem but gives no direction for how to address it. & Poor comment that is largely irrelevant to the paper's scope. & Minor textual edit, such as revising wording or rephrasing. \\
3 & Minor gap, such as a missing citation or small omission. & Offers only a vague hint, such as suggesting that other approaches be considered. & Weak comment that is valid but superficial or obvious. & Small addition, such as adding a citation or a clarifying sentence. \\
4 & Noticeable weakness, such as an incomplete explanation or unclear wording. & Provides a generic suggestion by naming a category of change without specifying implementation. & Below-average comment that is pedantic or low in substantive value. & Paragraph-level revision, such as expanding an explanation or adding a paragraph. \\
5 & Moderate concern, such as an ambiguous methodological detail or missing justification. & Gives partial direction by pointing to a specific element that might help. & Average comment that identifies a valid but relatively limited issue. & Section-level rewrite, such as substantially revising a section or rewriting the discussion. \\
\bottomrule
\end{tabular}
\end{table}

\begin{table}[p]
\centering
\small
\setlength{\tabcolsep}{4pt}
\caption{Score anchors used to extract the four continuous metrics (continued)}
\begin{tabular}{@{}p{0.06\textwidth}p{0.215\textwidth}p{0.215\textwidth}p{0.215\textwidth}p{0.215\textwidth}@{}}
\toprule
Score & Opinion strength & Constructiveness & Comment quality & Revision cost \\
\midrule
6 & Significant issue, such as a questionable design choice or weak evidence for a claim. & Gives clear direction by naming a concrete method, analysis, or revision path. & Above-average comment that correctly identifies a real scientific issue. & Figure or table redo, or a comparable non-trivial reconstruction of presented results. \\
7 & Major flaw, such as an inappropriate method or unsupported key conclusion. & Provides detailed guidance, including both a recommended action and the rationale for it. & Good comment that is insightful and improves rigor. & Partial reanalysis, such as rerunning analyses for a subset or adding a robustness check. \\
8 & Serious threat to the validity of the main findings. & Gives an actionable prescription that specifies implementation details in addition to the rationale. & Very good comment that identifies an important issue that others might miss. & Major reanalysis, such as reanalyzing the full dataset or redoing the full statistical workflow. \\
9 & Critical flaw that invalidates a major portion of the results. & Provides a comprehensive roadmap, often including multiple options, contingencies, or references. & Excellent comment that substantially improves the paper's inferential core. & New data collection or additional experiment beyond revision of existing material. \\
10 & Fatal problem, such as a fundamental flaw in the premise or a rejection-level concern. & Transformative mentorship that supplies a complete repair roadmap with theory, examples, and alternatives. & Exceptional, transformative observation that could materially reshape the paper. & Complete overhaul, such as redesigning the study or collecting an entirely new dataset. \\
\bottomrule
\end{tabular}
\end{table}

\begin{table}[t]
\centering
\small
\caption{Keyword patterns used for supplementary validation of the four continuous metrics}
\label{tab:si_keywords}
\begin{tabular}{@{}p{0.18\textwidth}p{0.72\textwidth}@{}}
\toprule
Metric & Keyword patterns \\
\midrule
Opinion strength & \texttt{reject}, \texttt{flaw}, \texttt{unsupported}, \texttt{defective}, \texttt{fatal}, \texttt{major concern}, \texttt{not convincing}, \texttt{cannot recommend}, \texttt{cannot support}, \texttt{cannot justify}, \texttt{serious concern}, \texttt{critical weakness}, \texttt{not valid} \\
Constructiveness & \texttt{specify}, \texttt{please clarify}, \texttt{please compare}, \texttt{should clarify}, \texttt{should be compared}, \texttt{please specify}, \texttt{should be specified}, \texttt{should be described}, \texttt{should be addressed}, \texttt{should be discussed}, \texttt{should be provided}, \texttt{should be shown}, \texttt{should be included}, \texttt{should be clarified}, \texttt{would be helpful to} \\
Comment quality & \texttt{confounding}, \texttt{mechanism}, \texttt{generalize}, \texttt{generalizability}, \texttt{reproduce}, \texttt{reproducibility}, \texttt{endogeneity}, \texttt{alternative explanation}, \texttt{causal claim}, \texttt{confounding factor}, \texttt{measurement error}, \texttt{robustness check}, \texttt{omitted variable} \\
Revision cost & \texttt{reanalyze}, \texttt{rebuild}, \texttt{recompute}, \texttt{additional experiments}, \texttt{new data}, \texttt{new experiments}, \texttt{major revision}, \texttt{conducted additional experiment}, \texttt{performed additional experiment}, \texttt{reanalyzed the data}, \texttt{collected new data} \\
\bottomrule
\end{tabular}
\end{table}

\begin{table}[t]
\centering
\small
\caption{Overall single-metric OLS results.}
\label{tab:overall_single_metric_ols}
\begin{tabular}{@{}p{0.28\textwidth}p{0.24\textwidth}cc@{}}
\toprule
Predictor & Outcome & Coef. & Sig. \\
\midrule
Opinion strength & $\log(1 + C_3)$ & 0.0383 & ** \\
Constructiveness & $\log(1 + C_3)$ & -0.0195 & \\
Comment quality & $\log(1 + C_3)$ & 0.0492 & ** \\
Revision cost & $\log(1 + C_3)$ & 0.0964 & *** \\
Persuasion success & $\log(1 + C_3)$ & -0.0879 & \\
Rebuttal rate & $\log(1 + C_3)$ & -0.0244 & \\
\bottomrule
\end{tabular}
\end{table}

\begin{table}[t]
\centering
\small
\caption{OLS results for opinion strength by opinion type.}
\label{tab:opinion_strength_by_type_ols}
\begin{tabular}{@{}p{0.30\textwidth}p{0.18\textwidth}p{0.18\textwidth}cc@{}}
\toprule
Opinion/feedback type & Predictor & Outcome & Coef. & Sig. \\
\midrule
Novelty/Contribution & Opinion strength & $\log(1 + C_3)$ & 0.0443 & *** \\
Analysis/Interpretation & Opinion strength & $\log(1 + C_3)$ & 0.0403 & ** \\
Conceptual & Opinion strength & $\log(1 + C_3)$ & 0.0314 & ** \\
Presentation & Opinion strength & $\log(1 + C_3)$ & 0.0212 & \\
Scope & Opinion strength & $\log(1 + C_3)$ & 0.0194 & \\
Logic & Opinion strength & $\log(1 + C_3)$ & 0.0116 & \\
Recommend Reference & Opinion strength & $\log(1 + C_3)$ & 0.0067 & \\
Methodology & Opinion strength & $\log(1 + C_3)$ & -0.0038 & \\
\bottomrule
\end{tabular}
\end{table}

\begin{table}[t]
\centering
\small
\caption{OLS results for constructiveness by opinion type.}
\label{tab:constructiveness_by_type_ols}
\begin{tabular}{@{}p{0.30\textwidth}p{0.18\textwidth}p{0.18\textwidth}cc@{}}
\toprule
Opinion/feedback type & Predictor & Outcome & Coef. & Sig. \\
\midrule
Methodology & Constructiveness & $\log(1 + C_3)$ & -0.0217 & \\
Analysis/Interpretation & Constructiveness & $\log(1 + C_3)$ & 0.0199 & \\
Conceptual & Constructiveness & $\log(1 + C_3)$ & -0.0180 & \\
Novelty/Contribution & Constructiveness & $\log(1 + C_3)$ & -0.0125 & \\
Logic & Constructiveness & $\log(1 + C_3)$ & 0.0162 & \\
Scope & Constructiveness & $\log(1 + C_3)$ & -0.0147 & \\
Recommend Reference & Constructiveness & $\log(1 + C_3)$ & -0.0118 & \\
Presentation & Constructiveness & $\log(1 + C_3)$ & -0.0061 & \\
\bottomrule
\end{tabular}
\end{table}

\begin{table}[t]
\centering
\small
\caption{OLS results for comment quality by opinion type.}
\label{tab:comment_quality_by_type_ols}
\begin{tabular}{@{}p{0.30\textwidth}p{0.18\textwidth}p{0.18\textwidth}cc@{}}
\toprule
Opinion/feedback type & Predictor & Outcome & Coef. & Sig. \\
\midrule
Analysis/Interpretation & Comment quality & $\log(1 + C_3)$ & 0.0807 & *** \\
Presentation & Comment quality & $\log(1 + C_3)$ & 0.0454 & ** \\
Conceptual & Comment quality & $\log(1 + C_3)$ & 0.0281 & \\
Logic & Comment quality & $\log(1 + C_3)$ & 0.0233 & \\
Recommend Reference & Comment quality & $\log(1 + C_3)$ & -0.0242 & \\
Novelty/Contribution & Comment quality & $\log(1 + C_3)$ & 0.0128 & \\
Methodology & Comment quality & $\log(1 + C_3)$ & 0.0126 & \\
Scope & Comment quality & $\log(1 + C_3)$ & 0.0044 & \\
\bottomrule
\end{tabular}
\end{table}

\begin{table}[t]
\centering
\small
\caption{OLS results for revision cost by feedback type.}
\label{tab:revision_cost_by_type_ols}
\begin{tabular}{@{}p{0.30\textwidth}p{0.18\textwidth}p{0.18\textwidth}cc@{}}
\toprule
Opinion/feedback type & Predictor & Outcome & Coef. & Sig. \\
\midrule
Accept \& Revise & Revision cost & $\log(1 + C_3)$ & 0.0677 & *** \\
Rebut/Disagree & Revision cost & $\log(1 + C_3)$ & 0.0414 & *** \\
Clarify Misunderstanding & Revision cost & $\log(1 + C_3)$ & 0.0186 & \\
Promise for Future Work & Revision cost & $\log(1 + C_3)$ & 0.0185 & \\
Partially Accept & Revision cost & $\log(1 + C_3)$ & -0.0015 & \\
\bottomrule
\end{tabular}
\end{table}

\clearpage

\begin{figure}[!htbp]
  \centering
  \includegraphics[width=\textwidth,height=0.72\textheight,keepaspectratio]{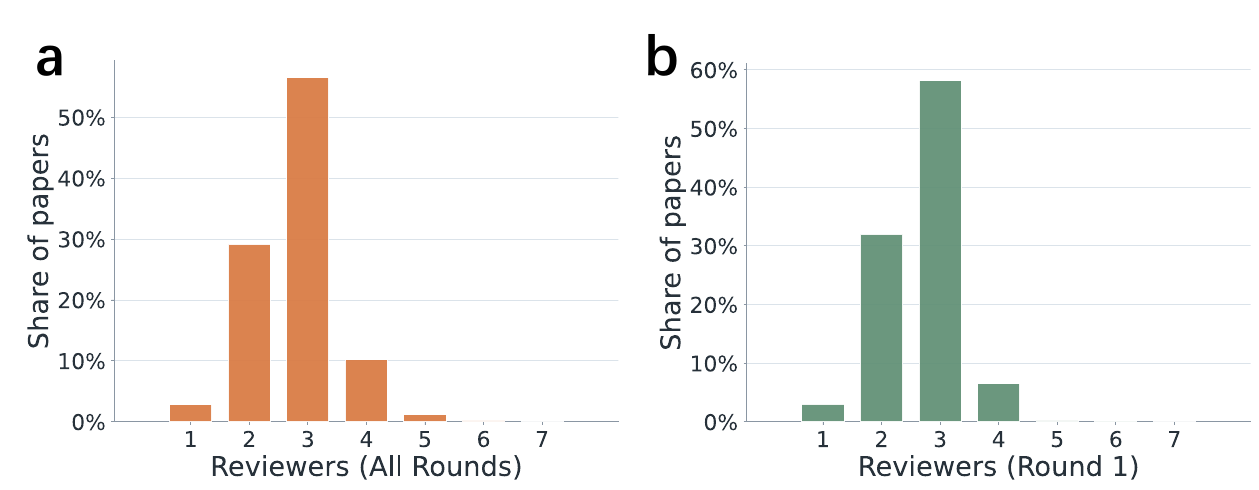}
  \caption{Supplementary Figure 1. Distributions of reviewer counts. (a) Distribution of the number of reviewers across all review rounds. (b) Distribution of the number of reviewers in the first review round.}
  \label{fig:sifig1}
\end{figure}

\begin{figure}[!htbp]
  \centering
  \includegraphics[width=\textwidth,height=0.72\textheight,keepaspectratio]{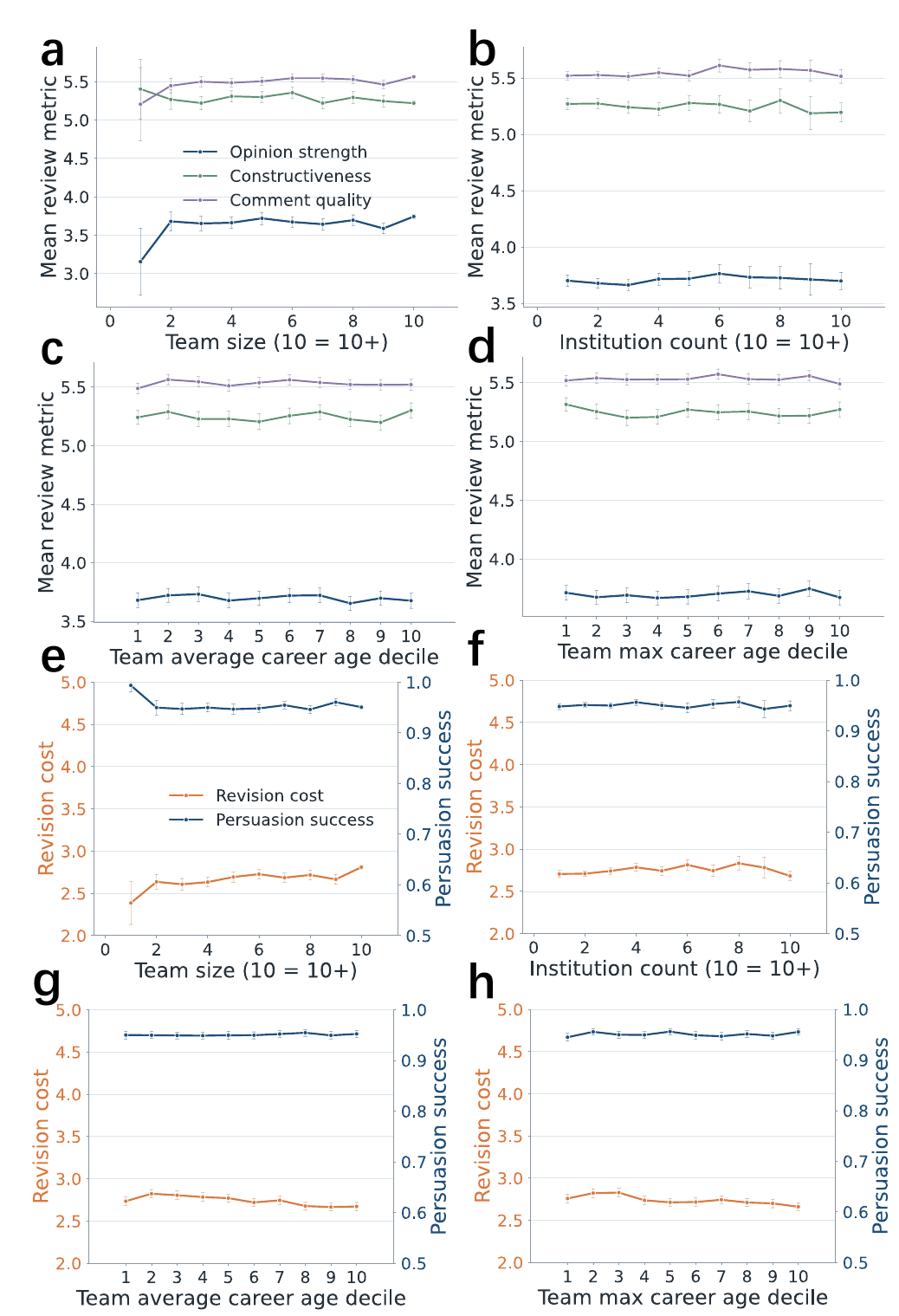}
  \caption{Supplementary Figure 2. Relationships between team characteristics and extracted peer-review metrics. This figure complements Table~1 in the main text by visualizing how team size, institution count, and career-age variables relate to the review and response indicators. Consistent with the main analysis, most associations are weak in magnitude, while revision cost shows the clearest and most selective pattern of association with team structure.}
  \label{fig:sifig2}
\end{figure}

\begin{figure}[!htbp]
  \centering
  \includegraphics[width=\textwidth,height=0.72\textheight,keepaspectratio]{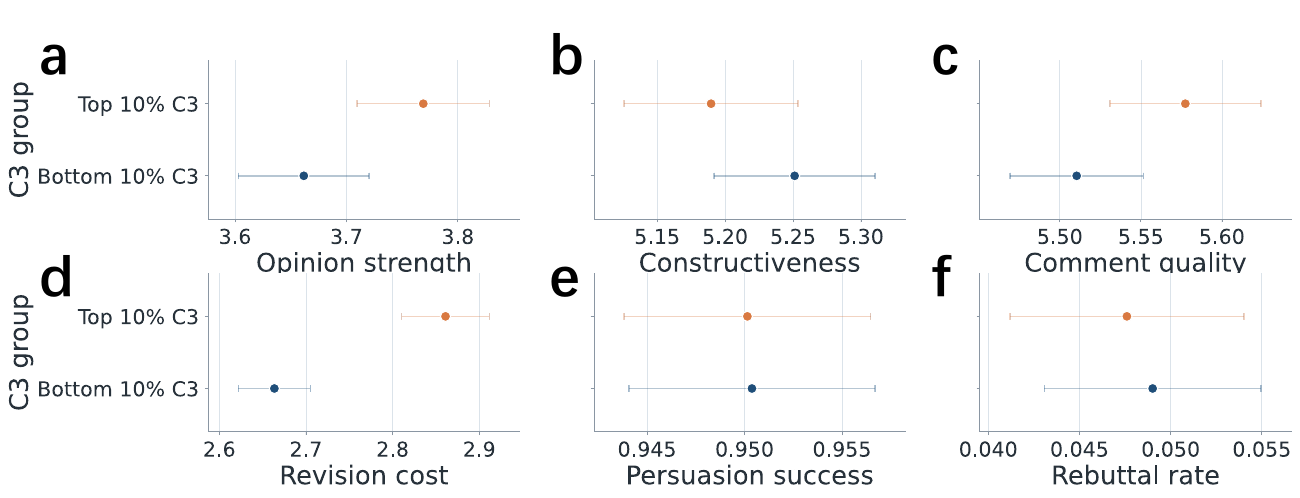}
  \caption{Supplementary Figure 3. Review and response profiles of papers in the top and bottom 10\% of $C_3$. Panels compare opinion strength, constructiveness, comment quality, revision cost, the issue-resolution proxy, and rebuttal rate between high- and low-impact papers. The figure summarizes the main impact-related pattern in the manuscript: within accepted papers, higher-impact papers are associated with stronger review, higher comment quality, and greater revision burden, whereas the issue-resolution proxy and rebuttal rate remain comparatively stable.}
  \label{fig:sifig3}
\end{figure}

\begin{figure}[!htbp]
  \centering
  \includegraphics[width=\textwidth,height=0.72\textheight,keepaspectratio]{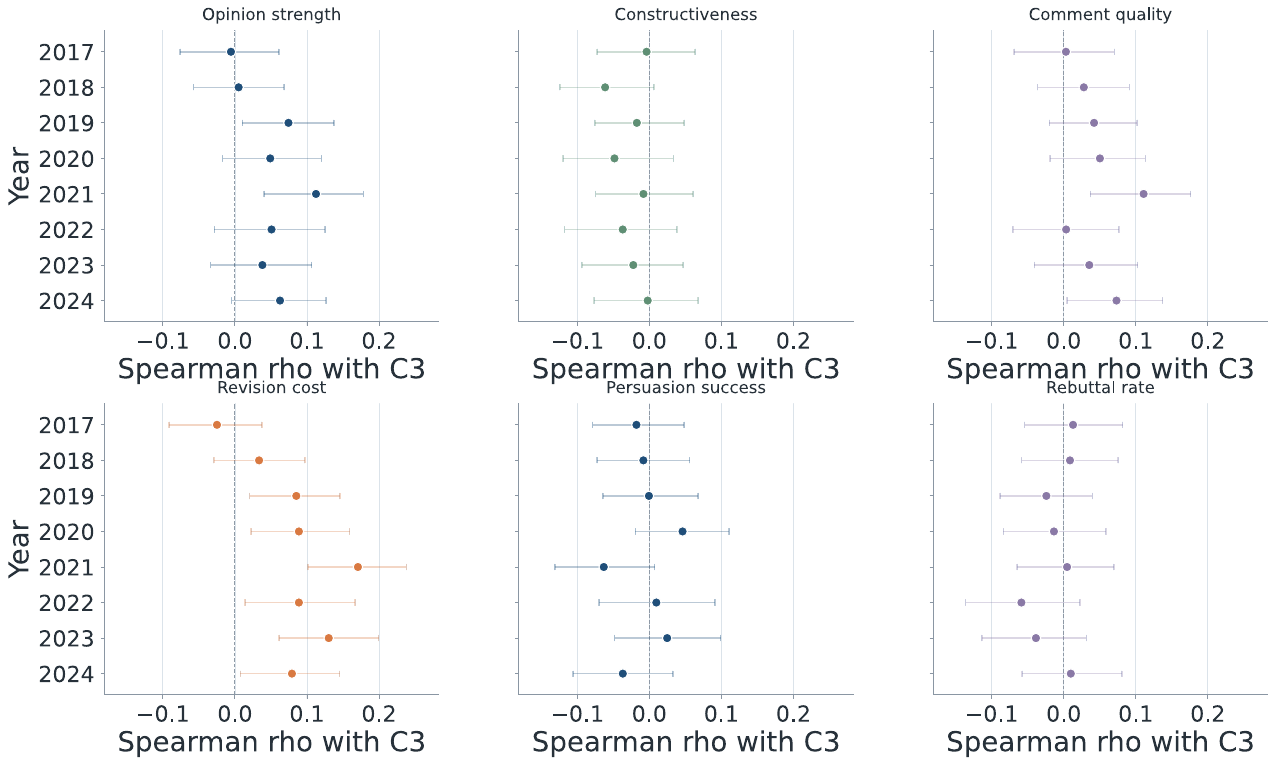}
  \caption{Supplementary Figure 4. Year-specific Spearman correlations between review/response metrics and $C_3$. Panels report yearly correlations for opinion strength, constructiveness, comment quality, revision cost, the issue-resolution proxy, and rebuttal rate. The figure shows that the most stable positive associations involve comment quality and revision cost, whereas constructiveness remains near zero or slightly negative and the issue-resolution proxy and rebuttal measures fluctuate around zero.}
  \label{fig:sifig4}
\end{figure}

\clearpage

\end{document}